\documentclass[11pt]{article}

\usepackage[final]{acl}
\usepackage{times}
\usepackage{latexsym}
\usepackage{subcaption}
\usepackage{array}
\usepackage{dblfloatfix}
\usepackage{tabularx}
\usepackage{subcaption}
\usepackage[justification=centering]{caption}

\usepackage[T1]{fontenc}

\usepackage[utf8]{inputenc}
\usepackage{booktabs}
\usepackage{multirow}
\usepackage{caption}
\usepackage{subcaption}
\usepackage{microtype}

\usepackage{inconsolata}

\usepackage{graphicx}
\usepackage{listings}

\lstdefinestyle{promptbox}{
  basicstyle=\ttfamily\small,
  frame=single,
  breaklines=true,
  columns=fullflexible,
  tabsize=2
}

\title{AffectDF: The Most Comprehensive Benchmark for Speech Deepfake Detection against Emotionally Expressive Attacks}

\author{
\textbf{Aurosweta Mahapatra\textsuperscript{1}},
\textbf{Xiutian Zhao\textsuperscript{1}},
\textbf{Shreeram Suresh Chandra\textsuperscript{1}},
\textbf{Zihan Zhang\textsuperscript{1}},
\\
\textbf{Zongyang Du\textsuperscript{1}},
\textbf{Ismail Rasim Ulgen\textsuperscript{1}},  
\textbf{Kong Aik Lee\textsuperscript{2}},
\textbf{Nicholas Andrews\textsuperscript{1}},\\ 
\textbf{Carlos Busso\textsuperscript{3}},
\textbf{Berrak Sisman\textsuperscript{1}}
\\[0.5em]
{\small
\textsuperscript{1} Johns Hopkins University \quad
\textsuperscript{2} Hong Kong Polytechnic University \quad
\textsuperscript{3} Carnegie Mellon University 
}
\\
{\small
\texttt{amahapa2@jhu.edu, noa@jhu.edu, busso@cmu.edu, sisman@jhu.edu}
}
}
\begin{document}
\maketitle

\begin{abstract}
Speech deepfake detection (SDD) systems achieve strong performance on conventional benchmarks; however, existing datasets provide limited coverage of emotionally expressive and recent large audio-language model (LALM)-based attacks. Existing emotional spoofing datasets are also limited in scale and attack diversity, typically covering only voice conversion (VC) or text-to-speech (TTS) attacks. We introduce AffectDF, the most comprehensive benchmark for emotionally expressive speech deepfakes, spanning TTS, VC, emotional VC, and LALM-based spoofing attacks across both acted and spontaneous emotional speech. AffectDF contains approximately 260 hours of speech generated using 21 spoofing attacks across five emotional states. We benchmark state-of-the-art SDD systems under conventional and emotional spoofing conditions, including LALM-based detectors evaluated with both inference-only prompting and supervised fine-tuning. Our experiments reveal severe robustness degradation when models trained on conventional benchmarks are evaluated on AffectDF, with several systems approaching near-random performance. Surprisingly, even large-scale emotional training does not consistently improve cross-domain robustness, indicating that current SDD systems fail to learn generalized spoof representations under emotional and prosodic variability. Robustness further varies substantially across emotional states, attack families, and acted vs spontaneous emotional speech conditions. These findings expose fundamental limitations of current SDD systems and establish AffectDF as a benchmark for developing more robust spoof detection models.

\end{abstract}

\section{Introduction}
Speech deepfake detection focuses on identifying synthetic speech generated using modern speech synthesis technologies such as text-to-speech and voice conversion \cite{SDD_survey, AS_Survey}. Recent advances in speech generation have enabled highly emotionally expressive synthetic speech capable of preserving speaker identity and modeling realistic prosody and emotional variability \cite{Emoqtts, F5TTS, StyleTTS2, Emo-ctrlTTS, expressive_vc, expressive_vc2}. As a result, the risk of malicious applications, including biometric spoofing, impersonation-based misinformation, and fraudulent voice calls, continues to grow \cite{misinfosurvey}. 

The development of SDD systems has been strongly driven by community challenges such as ASVspoof \cite{ASVspoof2015} and ADD \cite{ADD2022, ADD2023}, leading to widely used benchmarks including ASVspoof2019 \cite{ASVspoof2019} and ASVspoof5 \cite{ASVspoof2024}. These benchmarks enabled major advances in anti-spoofing architectures and evaluation under multilingual \cite{mlaad, multiling1, multiling2}, channel \cite{ASVspoof2021}, environmental \cite{envfake}, and in-the-wild spoofing conditions \cite{inthewild, for, spoofceleb, fakespeechwild}. Nevertheless, existing benchmarks primarily focus on neutral speech, while recent speech generation systems have become increasingly emotionally expressive.

Emotional speech introduces substantial variability in prosody, pitch, speaking rate, and temporal dynamics across different emotional states and speakers \cite{emotion, emotion3, prospkvariation}. Such variability can obscure spoof-related artifacts and alter acoustic patterns commonly exploited by SDD systems, making it difficult to learn generalized spoof cues \cite{emotionAS}.

Existing emotional spoofing datasets remain limited in scale and attack diversity. EmoFake \cite{EmoFake} primarily focuses on emotional VC attacks, while EmoSpoof-TTS \cite{emotionAS} focuses only on emotional TTS attacks with three spoofing systems. Furthermore, both datasets mainly rely on acted emotional speech from ESD \cite{ESD}, a widely used corpus in emotional speech synthesis, VC, and TTS research. As a result, understanding the role of acted emotional speech in SDD is important. Furthermore, acted versus spontaneous emotional speech \cite{Busso_202x, naturalvoices} remains largely unexplored in emotional spoofing literature. Existing emotional spoofing datasets also lack recent LALM-based attacks. These limitations make it difficult to understand how current SDD systems behave under diverse emotional conditions and modern spoofing scenarios. To better understand these challenges, we investigate the following research questions:


(i) \textbf{Robustness to Emotional Attacks:} Are current SDD systems trained on conventional benchmarks robust against emotionally expressive synthetic speech generated using diverse attack paradigms, including TTS, VC, EVC, and LALM-based systems?

(ii) \textbf{Cross-Domain Generalization:} Does training on large-scale emotional spoofing datasets improve generalization across both conventional and emotional spoofing benchmarks?

(iii) \textbf{Impact of Emotional Variability:} How does emotional variability affect SDD performance across emotional states, attack families, and acted versus spontaneous emotional speech conditions?

To address these challenges, we introduce AffectDF, which contains approximately 260 hours of speech generated using 21 spoofing attacks spanning TTS, VC, EVC, and LALM-based systems across five emotional states. The dataset includes both acted and spontaneous emotional speech, enabling evaluation under diverse emotional and recent spoofing conditions.  Using AffectDF, we benchmark multiple state-of-the-art SDD systems, including RawNet2~\cite{RawNet2}, AASIST~\cite{AASIST}, XLSR-SLS~\cite{ssl_sls}, XLSR-Mamba~\cite{xlsr-mamba}, ProSDD~\cite{prosdd}, and LALM-based detectors such as Qwen-2.5-Omni~\cite{qwen2.5omni}, Qwen-3.0-Omni~\cite{qwen3omni}, and Voxtral~\cite{liu2025voxtral}, evaluated under both inference-only and supervised fine-tuned settings. These SDD systems are evaluated across conventional and emotional spoofing benchmarks, including ASVspoof2019 \cite{ASVspoof2019}, ASVspoof5 \cite{ASVspoof2024}, EmoFake \cite{EmoFake}, and EmoSpoof-TTS \cite{emotionAS}. We analyze robustness, cross-domain generalization, and the impact of emotional variability on spoof detection performance.

Our main contributions are as follows: \textbf{(i)} we introduce \textbf{AffectDF, the most comprehensive benchmark for emotionally expressive speech deepfakes}, spanning TTS, VC, EVC, and LALM-based spoofing attacks across both acted and spontaneous emotional speech; \textbf{(ii)} we conduct a large-scale evaluation across conventional and emotional spoofing benchmarks, covering conventional and SSL-based SDD systems as well as LALM-based SDD under inference-only and supervised fine-tuning settings; \textbf{(iii)} we show that current SDD systems exhibit severe robustness degradation under emotionally expressive spoofing conditions; and \textbf{(iv)} we provide the first systematic analysis of SDD robustness across emotional states, attack families, and acted versus spontaneous emotional speech, showing that SDD systems fail to learn generalized spoof-relevant representations under emotional and prosodic variability.  \textbf{To the best of our knowledge, AffectDF is the first benchmark for systematically evaluating speech deepfake detection under diverse emotionally expressive spoofing attacks.} AffectDF and related resources are publicly available.\footnote{AffectDF: \url{https://affectdf33-data.github.io/AffectDF-Data/}}


\section{Related Work}
\paragraph{Speech Deepfake Detection.} Existing SDD systems include end-to-end architectures such as RawNet2 \cite{RawNet2} and AASIST \cite{AASIST}, as well as more recent SSL-based approaches \cite{ssl_rawnet2, ssl_sls, xlsr-mamba, prosdd, mahapatra2025hula}. More recently, LALM-based systems have also been explored for speech deepfake detection \cite{LLM-SDD, LLM-SDD2}. These systems are primarily developed and evaluated on conventional SDD benchmarks such as ASVspoof \cite{ASVspoof2015, ASVspoof2019, ASVspoof2024}, which focus on neutral speech and provide limited coverage of emotionally expressive spoofing conditions.


\paragraph{Emotionally Expressive Speech Deepfakes.} Recent work has explored emotionally expressive speech deepfakes through datasets such as EmoFake \cite{EmoFake} and EmoSpoof-TTS \cite{emotionAS}, showing that existing SDD systems are vulnerable to emotional synthetic speech attacks. However, compared to conventional SDD benchmarks, these datasets remain limited in scale, attack diversity, and generation paradigms. EmoFake contains seven emotional VC attacks, while EmoSpoof-TTS contains only three TTS attacks. Furthermore, both datasets mainly rely on acted emotional speech generated from ESD \cite{ESD}. Consequently, comprehensive robustness and cross-domain generalization analysis under emotionally expressive spoofing conditions remains difficult.

\paragraph{Research Gap.} The role of emotion in SDD remains poorly understood. Existing emotional spoofing datasets are limited in scale, emotional diversity, and attack coverage. As a result, it remains unclear whether current SDD systems can generalize across diverse emotional conditions and modern expressive spoofing attacks. We address these limitations through a comprehensive benchmark for emotional speech deepfake detection.


\begin{table*}[t]
\centering
\small
\setlength{\tabcolsep}{5pt}
\renewcommand{\arraystretch}{1.15}
\caption{
Dataset composition and split configuration of AffectDF across train, development, and test partitions.
}
\begin{tabular}{ccccccc}
\toprule

\textbf{Attack ID} &
\textbf{Generation Type} &
\textbf{Models} &
\textbf{Base Data} &
\textbf{Samples} &
\textbf{No. Spks} &
\textbf{Split} \\

\midrule

\multicolumn{7}{c}{\textbf{TRAIN}} \\
\midrule

A01 &
LALM-EVC &
Qwen 2.5-Omni &
ESD &
35,000 &
4 &
Train \\

A02 &
LALM-EVC &
Qwen 2.5-Omni (steered) &
ESD &
28,000 &
4 &
Train \\

A03 &
TTS &
CosyVoice &
ESD &
5,999 &
4 &
Train \\

A04 &
TTS &
CosyVoice2 &
ESD &
6,000 &
4 &
Train \\

A05 &
TTS &
Qwen3-TTS &
ESD &
6,000 &
4 &
Train \\

\midrule

\multicolumn{7}{c}{\textbf{DEVELOPMENT}} \\
\midrule

A06 &
LALM-EVC &
MiniCPM &
ESD &
17,500 &
2 &
Dev \\

A07 &
TTS &
CosyVoice3 &
ESD &
2830 &
2 &
Dev \\

\midrule

\multicolumn{7}{c}{\textbf{TEST}} \\
\midrule

A08 &
LALM-EVC &
Kimi-Audio &
ESD &
34,992 &
4 &
Test \\

A09 &
LALM-EVC &
Kimi-Audio (steered) &
ESD &
27,994 &
4 &
Test \\

A10 &
VC+EVC &
Vevo2 &
ESD &
24,000 &
4 &
Test \\

A11 &
VC &
Vevo 2 &
ESD &
6,000 &
4 &
Test \\

A12 &
EVC &
GenVC &
ESD &
24,000 &
4 &
Test \\

A13 &
VC &
GenVC &
ESD &
6,000 &
4 &
Test \\

A14 &
VC &
GenVC &
MSP &
4108 &
4 &
Test \\

A15 &
VC &
ConsistencyVC &
ESD &
6,052 &
4 &
Test \\

A16 &
VC &
TriAAN-VC &
ESD &
6,000 &
4 &
Test \\

A17 &
VC &
DDDMVC &
ESD &
6,000 &
4 &
Test \\

A18 &
TTS &
Style-TTS2 &
ESD &
6,000 &
4 &
Test \\

A19 &
TTS &
Style-TTS2 &
MSP &
4107 &
4 &
Test \\

A20 &
TTS &
F5-TTS &
ESD &
6,000 &
4 &
Test \\

A21 &
TTS &
F5-TTS &
MSP &
4107 &
4 &
Test \\

\bottomrule
\end{tabular}
\label{tab:dataset_split}
\end{table*}

\section{The AffectDF Benchmark}
\subsection{Design Motivation}

Recent advances in expressive speech generation have enabled highly natural emotional speech synthesis using TTS, VC, EVC, and LALM-based systems. These systems jointly model speaker identity, prosody, and emotional expression, producing increasingly realistic synthetic speech \cite{F0-2,F0-3,F0_vc,F0_vc2}. Unlike neutral speech, emotional speech introduces substantial variability in pitch, energy, speaking rate, shimmer, and temporal dynamics across emotional states and speakers, potentially altering spoof-related acoustic patterns exploited by current SDD systems. Furthermore, real-world emotional expression is often spontaneous rather than acted \cite{naturalvoices-Ali, Busso_202x, naturalvoices}, exhibiting substantially different acoustic and prosodic characteristics. Motivated by these challenges, AffectDF spans TTS, VC, EVC, and LALM-based attacks across both acted and spontaneous emotional speech, enabling systematic evaluation of emotional robustness, cross-domain generalization, and emotional variability in speech deepfake detection.

\subsection{Dataset Construction}
\subsubsection{Source Corpora and Emotional Conditions}
AffectDF is constructed using both acted and spontaneous emotional speech corpora to enable evaluation across diverse speaking styles and emotional conditions. We use the Emotional Speech Dataset (ESD) \cite{ESD} as the primary acted emotional speech corpus and the MSP-Podcast \cite{Busso_202x} corpus as the spontaneous emotional speech database. The ESD corpus contains professionally acted emotional speech with parallel utterances across multiple emotional states, while the MSP-Podcast corpus contains naturally occurring spontaneous emotional speech with substantially different prosodic and acoustic characteristics. AffectDF includes five emotional states: \textit{neutral, happiness, anger, sadness, and surprise}. By combining acted and spontaneous emotional speech, AffectDF enables controlled analysis of how speaking style and emotional variability affect speech deepfake detection performance.
\begin{figure}[t]
    \centering
    \includegraphics[width=\columnwidth,height=0.25\textheight]{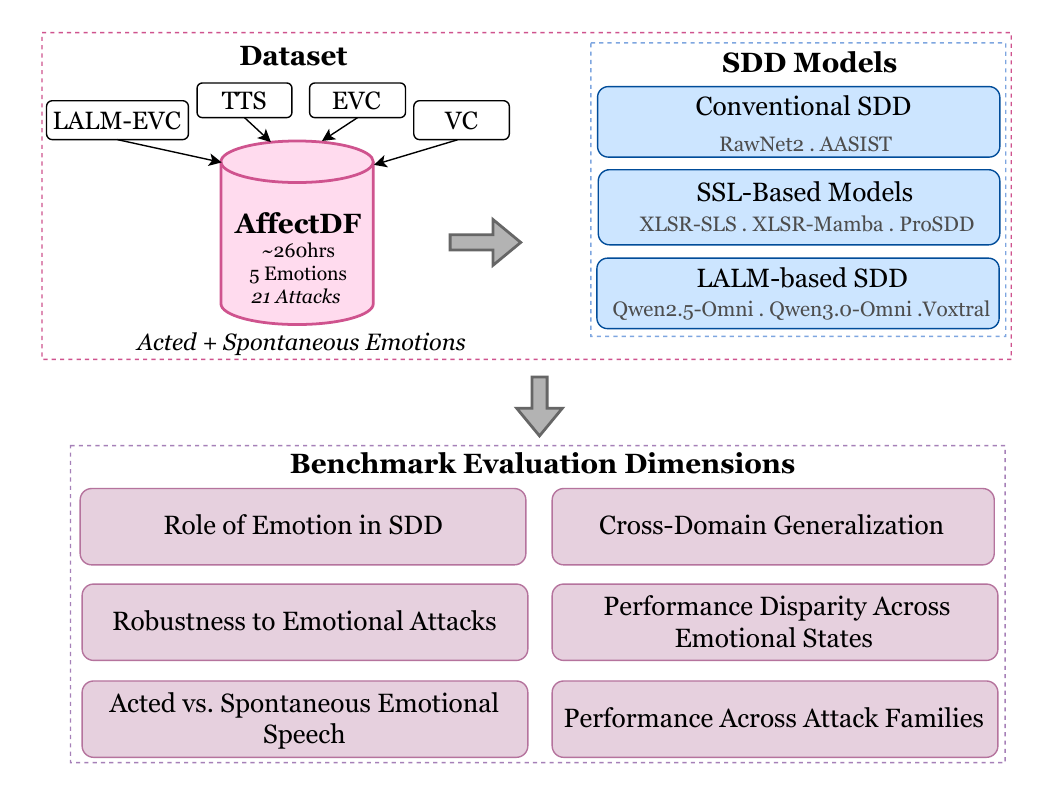}
    \caption{Overview of AffectDF benchmark, including the proposed dataset, evaluated SDD models, and evaluation dimensions.}
    \label{fig:overview}
\end{figure}
\subsubsection{Spoofing Attack Taxonomy}
AffectDF contains 21 spoofing attacks spanning TTS, VC, EVC, VC+EVC, and LALM-based speech generation systems. The dataset includes multiple recent emotional speech generation models, enabling evaluation across diverse synthesis paradigms and attack conditions. The attack framework includes recent systems such as (i) \textbf{TTS}: CosyVoice \cite{Cosyvoice}, CosyVoice2 \cite{cosyvoice2}, CosyVoice3 \cite{cosyvoice3}, StyleTTS2 \cite{StyleTTS2}, F5-TTS \cite{F5TTS}, and Qwen3-TTS \cite{qwen3tts} (ii) \textbf{VC}: GenVC \cite{genvc}, ConsistencyVC \cite{consistencyVC}, TriAAN-VC \cite{TriAAN-VC}, and DDDMVC \cite{dddmvc} \cite{genvc,dddmvc}, (iii) \textbf{EVC}: GenVC and Vevo2 \cite{genvc,vevo2}, and (iv) \textbf{LALM-based EVC}:  Qwen2.5-Omni \cite{qwen2.5omni}, Kimi-Audio \cite{kimi}, and MiniCPM \cite{minicpm}. Detailed attack configurations and dataset statistics are summarized in Table~\ref{tab:dataset_split} and Appendix \ref{sec:affectDF_deets}.

\subsubsection{Train, Development, and Test Splits}

The train, development, and test partitions are designed with disjoint speakers and attack systems to evaluate cross-speaker and cross-attack generalization. Overall split statistics are summarized in Table~\ref{tab:dataset_split} and Appendix Tables~\ref{tab:dataset_comparison}, ~\ref{tab:affectdf-split-emotion-counts}, and ~\ref{tab:dataset-statistics}.

\paragraph{Train.} The training split contains five attacks \textit{(A01--A05)} generated from four speakers using LALM-based EVC, LALM-based TTS, and conventional TTS systems. The split contains 86,999 samples, including 80,999 spoofed and 6,000 bona fide samples. 

\paragraph{Development.} The development split contains two attacks \textit{(A06--A07)} generated using LALM-based EVC and conventional TTS systems. It contains 23,330 samples, including 20,330 spoofed and 3,000 bona fide samples.

\paragraph{Test.} The test partition is designed to be more challenging along multiple dimensions. It contains 14 attacks \textit{(A08--A21)} spanning TTS, VC, EVC, and LALM-EVC systems. The test set uses acted emotional speech from ESD and spontaneous emotional speech from MSP-Podcast as base data, enabling evaluation across both settings. The test set contains eight speakers in total, four from ESD and four from MSP-Podcast, and includes 175,468 samples, with 165,360 spoofed samples and 10,108 bona fide samples. By increasing speaker diversity, introducing VC and non-LALM EVC attacks, and incorporating spontaneous emotional speech, the test partition enables a challenging evaluation of whether SDD systems trained on acted emotional synthetic speech can generalize to diverse emotional spoofing conditions.

\subsection{Acted vs. Spontaneous Emotional Speech}

AffectDF is designed to enable controlled comparison between acted and spontaneous emotional speech. We use generation models, GenVC \cite{genvc}, StyleTTS2 \cite{StyleTTS2}, and F5-TTS \cite{F5TTS}, to generate both acted and spontaneous speech conditions. For these models, the acted speech attacks \textit{(A13, A18, A20)} use ESD as the base corpus, while the spontaneous speech attacks \textit{(A14, A19, A21)} use MSP-Podcast as the base corpus. Both conditions contain four speakers with disjoint speaker identities across the acted and spontaneous subsets. This design allows analysis of how spoof detection performance changes when the generation method is kept fixed but the underlying speech style changes from acted to spontaneous. 


\subsection{Benchmark Protocol and Analysis Capabilities}
We release detailed protocol files containing speaker ID, audio ID, attack ID, emotion label, generation method, generation model, and spoof/real label. These metadata allow the research community to go beyond binary real-versus-spoof evaluation and analyze how emotion, attack type, generation model, and speech style (acted/spontaneous) affect detection performance. AffectDF follows a controlled, leave-one-factor-varied design, where many utterances share parallel source content or speaker conditions while differing in one key factor, such as synthesis method, emotional state, or base speech style. This structure enables targeted analysis of how individual factors influence spoof detection performance and supports more interpretable evaluation of SDD systems.

\subsection{Comparison with ASVspoof5}

ASVspoof5 \cite{ASVspoof2024} is one of the largest recent SDD benchmarks, containing diverse spoofing conditions including TTS, VC, and adversarial attacks. However, it does not provide emotion labels or support controlled evaluation across emotional states and speaking styles. In contrast, AffectDF focuses on emotionally expressive speech deepfakes spanning TTS, VC, EVC, and LALM-based attacks across acted and spontaneous emotional speech. SER experiments reported in Appendix~\ref{sec:appendix_SER} further show that AffectDF exhibits substantially stronger emotional characteristics than existing SDD benchmarks. Furthermore, AffectDF provides metadata including emotion labels, attack type, generation model, speaker ID, and source corpus information, enabling fine-grained analysis of emotional robustness and cross-domain generalization.

\begin{table*}[t]
\centering
\small
\caption{
Cross-domain evaluation of SDD models trained on ASVspoof2019, ASVspoof5, and AffectDF.
}
\setlength{\tabcolsep}{4pt}
\renewcommand{\arraystretch}{1.2}
\begin{tabular}{llccccc}
\toprule
\textbf{Model} & \textbf{Train Data}
& \textbf{ASVspoof-2019}
& \textbf{ASVspoof-2021}
& \textbf{ASVspoof5}
& \textbf{EmoFake}
& \textbf{AffectDF} \\
\midrule
RawNet2     & \textit{ASVspoof2019} & 4.60  & 8.08  & 40.67 & 21.71 & 59.71 \\
AASIST      & \textit{ASVspoof2019} & 0.83  & 8.15  & 35.53 & 13.64 & 56.40 \\
XLSR-SLS    & \textit{ASVspoof2019} & 0.56  & 3.04  & 25.43 & 8.84  & 44.91 \\
XLSR-Mamba  & \textit{ASVspoof2019} & 0.20  & 1.64  & 15.54 & 0.69  & 29.78 \\
ProSDD      & \textit{ASVspoof2019} & 0.42  & 3.87  & 16.14 & 3.70  & 31.04 \\
\midrule
RawNet2     & \textit{ASVspoof5} & 24.75 & 25.59 & 43.61 & 49.49 & 33.02 \\
AASIST      & \textit{ASVspoof5} & 23.16 & 22.74 & 25.77 & 62.71 & 18.00 \\
XLSR-SLS    & \textit{ASVspoof5} & 27.00 & 26.54 & 39.62 & 58.57 & 32.28 \\
XLSR-Mamba  & \textit{ASVspoof5} & 13.65 & 13.67 & 7.27  & 20.77 & 35.27 \\
ProSDD      & \textit{ASVspoof5} & 19.04 & 18.08 & 7.38  & 25.06 & 12.49 \\
\midrule
RawNet2     & \textit{AffectDF} & 43.02 & 47.25 & 44.65 & 6.59  & 23.17 \\
AASIST      & \textit{AffectDF} & 44.52 & 48.78 & 45.75 & 22.60 & 36.91 \\
XLSR-SLS    & \textit{AffectDF} & 64.83 & 60.56 & 56.54 & 18.41 & 36.20 \\
XLSR-Mamba  & \textit{AffectDF} & 46.47 & 53.09 & 47.67 & 28.39 & 41.60 \\
ProSDD      & \textit{AffectDF} & 58.15 & 56.43 & 63.16 & 22.08 & 48.14 \\
\bottomrule
\end{tabular}
\vspace{-1mm}
\label{tab:main_results}
\end{table*}

\section{Experimental Design}

\paragraph{Benchmark Evaluation Dimensions.} As shown in Figure \ref{fig:overview}, AffectDF supports evaluation across six benchmark dimensions: (i) role of emotion in SDD, (ii) cross-domain generalization, (iii) robustness to emotional attacks, (iv) performance disparity across emotional states, (v) acted vs. spontaneous emotional speech, and (vi) performance across attack families. We use Equal Error Rate (EER) as the primary evaluation metric, and report EER (\%) following standard practice in speech deepfake detection \cite{ASVspoof2019}. We additionally report min t-DCF results in Appendix~\ref{sec:tDCF}. To analyze the emotional characteristics of AffectDF and existing SDD benchmarks, we conduct Speech Emotion Recognition (SER) experiments, reported in Appendix~\ref{sec:appendix_SER}. Results across these benchmark dimensions are presented in Section~\ref{sec:results}.


\paragraph{Datasets.} We evaluate SDD systems on both conventional and emotional spoofing benchmarks, including ASVspoof2019 \cite{ASVspoof2019}, ASVspoof2021 \cite{ASVspoof2021}, ASVspoof5 \cite{ASVspoof2024}, EmoFake \cite{EmoFake}, and EmoSpoof-TTS \cite{emotionAS}. We use the official train/development/evaluation partitions for ASVspoof2019 and ASVspoof5, the evaluation partition of ASVspoof2021, the English evaluation partition of EmoFake, and the test partition of EmoSpoof-TTS. Additional dataset details are provided in Appendix~\ref{sec:appendix_datasets}.

\paragraph{Baseline Models.} We evaluate conventional, SSL-based, and LALM-based SDD systems. Our baselines include RawNet2 \cite{RawNet2}, AASIST \cite{AASIST}, XLSR-SLS \cite{ssl_sls}, XLSR-Mamba \cite{xlsr-mamba}, and ProSDD \cite{prosdd}. For LALM-based evaluation, we use Qwen-2.5-Omni~\cite{qwen2.5omni}, Qwen-3.0-Omni~\cite{qwen3omni}, and Voxtral~\cite{liu2025voxtral}. We evaluate these models under prompt-based inference settings and further fine-tune Voxtral to assess supervised LALM-based SDD. We follow the original training configurations for all trainable baselines. Additional model and reproducibility details are provided in Appendices~\ref{sec:appendix_baselines} and~\ref{sec:model_training_details}. LALM inference and fine-tuning details are provided in Appendices~\ref{sec:appendix_prompts_detect} and~\ref{sec:finetuning_details}.

\section{Results}
\label{sec:results}
\subsection{Emotion Breaks Cross-Domain Generalization}
\begin{table*}[t]
\centering
\small
\caption{Emotion-wise EER (\%) results. Neutral (N), Happy (H), Angry (A), Sad (S), Surprise (Su).}
\setlength{\tabcolsep}{4pt}
\renewcommand{\arraystretch}{1.2}
\begin{tabular}{llccccccccccc}
\toprule
\multirow{2}{*}{\textbf{Model}} & \multirow{2}{*}{\textbf{Train Data}}
& \multicolumn{5}{c}{\textbf{EmoFake}}
& \multicolumn{5}{c}{\textbf{AffectDF}} \\
\cmidrule(lr){3-7} \cmidrule(lr){8-12}
& & N & H & A & S & Su & N & H & A & S & Su \\
\midrule
RawNet2    & \textit{ASVspoof2019} & 19.66 & 16.57 & 23.00 & -- & 28.54 & 56.21 & 60.93 & 63.82 & 48.88 & 64.28 \\
AASIST     & \textit{ASVspoof2019} & 17.20 & 11.63 & 15.00 & -- & 13.57 & 65.45 & 61.07 & 50.91 & 41.82 & 48.56 \\
XLSR-SLS   & \textit{ASVspoof2019} & 5.20  & 8.09  & 9.03  & -- & 10.77 & 59.94 & 51.92 & 31.87 & 32.12 & 39.46 \\
XLSR-Mamba & \textit{ASVspoof2019} & 0.86  & 0.43  & 0.94  & -- & 0.54  & 28.52 & 31.78 & 26.13 & 21.42 & 23.99 \\
ProSDD     & \textit{ASVspoof2019} & 2.14  & 3.60  & 4.57  & -- & 2.29  & 31.70 & 33.56 & 26.20 & 24.08 & 33.14 \\
Voxtral    & \textit{ASVspoof2019} & 5.29  & 8.24  & 8.24  & -- & 9.71   & 31.52 & 28.25 & 19.99 & 16.92 & 21.85 \\
\midrule
RawNet2    & \textit{ASVspoof5} & 52.29 & 48.07 & 42.86 & -- & 45.30 & 42.25 & 33.99 & 25.00 & 30.44 & 20.38 \\
AASIST     & \textit{ASVspoof5} & 70.46 & 59.51 & 56.34 & -- & 63.06 & 22.48 & 17.48 & 14.04 & 22.00 & 10.98 \\
XLSR-SLS   & \textit{ASVspoof5} & 59.71 & 55.67 & 59.16 & -- & 58.50 & 39.73 & 30.74 & 27.05 & 38.49 & 20.25 \\
XLSR-Mamba & \textit{ASVspoof5} & 20.71 & 17.29 & 17.00 & -- & 22.29 & 40.85 & 41.55 & 22.83 & 23.14 & 20.06 \\
ProSDD     & \textit{ASVspoof5} & 27.46 & 22.17 & 20.94 & -- & 18.74 & 16.25 & 11.95 & 9.60  & 13.83 & 7.35  \\
\midrule
RawNet2    & \textit{AffectDF} & 7.51  & 5.00  & 6.66  & -- & 5.14  & 25.67 & 26.79 & 17.18 & 15.40 & 15.05 \\
AASIST     & \textit{AffectDF} & 25.06 & 23.14 & 26.71 & -- & 15.83 & 45.98 & 49.23 & 22.57 & 25.11 & 18.03 \\
XLSR-SLS   & \textit{AffectDF} & 14.86 & 16.57 & 15.63 & -- & 23.34 & 41.86 & 43.30 & 25.07 & 25.40 & 21.15 \\
XLSR-Mamba & \textit{AffectDF} & 29.00 & 31.26 & 30.66 & -- & 22.14 & 57.60 & 52.31 & 24.65 & 27.15 & 22.33 \\
ProSDD     & \textit{AffectDF} & 21.34 & 14.66 & 14.46 & -- & 21.20 & 46.11 & 50.02 & 43.84 & 48.03 & 35.23 \\
Voxtral    & \textit{AffectDF} & 32.80  & 28.87 & 32.94  & -- & 20.38  & 28.93 & 26.29 & 19.76 & 14.59 & 15.33 \\
\midrule
Qwen-2.5-Omni & \textit{Inference-only} & 29.37 & 22.96 & 25.50 & -- & 31.49 & 54.16 & 48.74 & 37.81 & 23.76 & 41.61 \\
Qwen-3.0-Omni & \textit{Inference-only} & 35.12 & 31.73 & 39.58 & -- & 39.74 & 41.14 & 37.69 & 42.47 & 40.85 & 44.99 \\
Voxtral        & \textit{Inference-only} & 52.24 & 50.81 & 52.79 & -- & 50.67 & 56.46 & 56.48 & 51.99 & 50.57 & 50.56 \\
\bottomrule
\end{tabular}
\label{tab:emotion_results}
\end{table*}

\paragraph{Training on ASVspoof2019.} As shown in Table~\ref{tab:main_results}, models trained on ASVspoof2019 perform strongly on conventional evaluation conditions such as ASVspoof2019 and ASVspoof2021, particularly SSL-based systems including XLSR-SLS, XLSR-Mamba, and ProSDD. However, performance degrades substantially on ASVspoof5 and emotional spoofing datasets. RawNet2 and AASIST exhibit severe degradation on AffectDF, reaching 59.71\% and 56.40\% EER, respectively. Although SSL-based systems perform comparatively better, robustness remains limited even for strong models such as XLSR-Mamba and ProSDD. Interestingly, several models achieve relatively low EERs on EmoFake while still failing on AffectDF, suggesting that robustness on smaller emotional datasets does not necessarily transfer to broader emotional spoofing conditions with higher attack and speaking-style diversity.

\paragraph{Training on ASVspoof5.} When trained on ASVspoof5, XLSR-Mamba and ProSDD generalize more effectively than RawNet2, AASIST, and XLSR-SLS under the challenging in-domain conditions of ASVspoof5. However, RawNet2, AASIST, and XLSR-SLS nearly collapse on EmoFake, which consists entirely of emotional VC attacks, while ASVspoof5 training is TTS-based. Although XLSR-Mamba and ProSDD perform comparatively better, robustness under emotional spoofing conditions remains limited. In contrast, performance on AffectDF improves noticeably for several systems, particularly AASIST and ProSDD. Compared to EmoFake, AffectDF contains a broader mixture of TTS, VC, EVC, and LALM-based attacks, enabling partial transfer of spoof-relevant cues learned from the more diverse spoofing variability present in ASVspoof5. Among all evaluated systems, ProSDD achieves the strongest performance on AffectDF with an EER of 12.49\%.

\paragraph{Training on AffectDF.} Training on AffectDF produces substantially different behavior across architectures. Performance on conventional benchmarks degrades sharply across nearly all models, indicating limited transferability from emotional synthetic speech to conventional spoofing conditions. Surprisingly, RawNet2, despite being the simplest evaluated architecture, achieves the strongest performance on AffectDF itself. At the same time, AffectDF training improves performance on EmoFake for RawNet2, AASIST, and XLSR-SLS compared with ASVspoof5 training, suggesting partial transfer of emotional spoof-relevant cues across emotional datasets. However, the gains remain inconsistent across architectures. Models such as XLSR-Mamba and ProSDD, which generalize strongly under conventional training conditions, degrade substantially when trained on AffectDF. These results suggest that large-scale emotional training does not consistently improve robustness and may substantially reduce generalization to conventional spoofing benchmarks. Additional EmoSpoof-TTS results are reported in Appendix~\ref{sec:emospooftts}.

\begin{table*}[t]
\centering
\small
\caption{
Evaluation of LALM-based SDD models under inference-only and fine-tuning settings.
}
\setlength{\tabcolsep}{5pt}
\renewcommand{\arraystretch}{1.2}
\begin{tabular}{llcccc}
\toprule
\textbf{Model} & \textbf{Train Data} & \textbf{ASVspoof2019} & \textbf{ASVspoof5} & \textbf{EmoFake} & \textbf{AffectDF} \\
\midrule
Qwen-2.5-Omni & \textit{Inference-only} & 29.54 & 46.23 & 25.15 & 45.29 \\
Qwen-3.0-Omni & \textit{Inference-only} & 42.42 & 27.91 & 34.18 & 39.81 \\
Voxtral & \textit{Inference-only} & 46.06 & 50.62 & 51.76 & 54.20 \\
Voxtral & \textit{ASVspoof2019} & 3.05 & 20.67 & 8.73 & 26.15 \\
Voxtral & \textit{AffectDF} & 19.52 & 56.54 & 28.96 & 26.09 \\
\bottomrule
\end{tabular}
\vspace{-1mm}
\label{tab:lalm_results}
\end{table*}

\vspace{0.5mm}
\noindent\textbf{Is the Bottleneck the Data or the Model?}
The results suggest that the limitations of current SDD systems cannot be explained solely by insufficient emotional training data. While AffectDF substantially increases emotional attack diversity and scale compared to prior emotional spoofing datasets, training on AffectDF does not consistently improve robustness across architectures or datasets. In many cases, improved performance on emotional spoofing conditions comes at the cost of severe degradation on conventional benchmarks. These results suggest that current SDD systems struggle to learn spoof-relevant representations that transfer consistently across emotional conditions, attack families, and speaking styles. Emotional speech introduces large variability in prosody, speaking rate, energy, and temporal dynamics, while different spoofing paradigms produce distinct synthesis artifacts. As a result, existing SDD models appear to rely heavily on domain- and attack-specific cues rather than generalized spoof representations.



\begin{table*}[!t]
\centering
\small
\setlength{\tabcolsep}{5pt}
\renewcommand{\arraystretch}{1.2}
\caption{
Attack-wise EER (\%) for acted and spontaneous emotional speech generation systems. Overall columns report pooled EER across attacks within each category.
}
\begin{tabular}{llcccccccc}
\toprule
\multirow{2}{*}{\textbf{Model}} &
\multirow{2}{*}{\textbf{Train Data}} &
\multicolumn{4}{c}{\textbf{Acted}} &
\multicolumn{4}{c}{\textbf{Spontaneous}} \\
\cmidrule(lr){3-6}
\cmidrule(lr){7-10}
& & \textbf{A13} & \textbf{A18} & \textbf{A20} & \textbf{Overall}
  & \textbf{A14} & \textbf{A19} & \textbf{A21} & \textbf{Overall} \\
\midrule
RawNet2    & \textit{ASVspoof2019} & 62.33 & 53.02 & 51.16 & 55.51 & 44.26 & 44.27 & 39.74 & 42.74 \\
AASIST     & \textit{ASVspoof2019} & 62.90 & 49.48 & 60.33 & 57.47 & 43.11 & 62.31 & 48.23 & 51.38 \\
XLSR-SLS   & \textit{ASVspoof2019} & 42.81 & 45.80 & 74.85 & 48.64 & 34.10 & 44.01 & 40.93 & 41.83 \\
XLSR-Mamba & \textit{ASVspoof2019} & 6.43  & 27.42 & 53.70 & 31.43 & 22.18 & 27.05 & 25.57 & 25.73 \\
ProSDD     & \textit{ASVspoof2019} & 14.22 & 18.47 & 15.57 & 16.46 & 9.43  & 54.15 & 22.49 & 29.72 \\
Voxtral     & \textit{ASVspoof2019} & 11.03 & 12.47 & 27.21 & 17.48 & 24.93  & 40.77 & 27.67 & 45.44 \\
\midrule
RawNet2    & \textit{ASVspoof5} & 35.25 & 36.40 & 41.07 & 37.25 & 23.27 & 53.01 & 62.36 & 47.48 \\
AASIST     & \textit{ASVspoof5} & 21.57 & 14.67 & 38.56 & 25.17 & 12.24 & 22.04 & 39.81 & 25.69 \\
XLSR-SLS   & \textit{ASVspoof5} & 36.93 & 25.85 & 41.70 & 34.90 & 32.62 & 32.82 & 62.82 & 42.66 \\
XLSR-Mamba & \textit{ASVspoof5} & 17.22 & 41.30 & 35.04 & 34.11 & 2.02  & 35.41 & 37.03 & 31.91 \\
ProSDD     & \textit{ASVspoof5} & 5.25  & 13.85 & 18.50 & 13.72 & 5.84  & 19.96 & 37.11 & 23.11 \\
\midrule
RawNet2    & \textit{AffectDF} & 34.38 & 1.58  & 1.35  & 18.97 & 4.79  & 2.63  & 2.69  & 3.64  \\
AASIST     & \textit{AffectDF} & 40.95 & 33.59 & 21.41 & 34.38 & 21.21 & 20.62 & 17.80 & 19.71 \\
XLSR-SLS   & \textit{AffectDF} & 30.45 & 28.95 & 23.97 & 28.28 & 17.06 & 26.25 & 27.44 & 24.67 \\
XLSR-Mamba & \textit{AffectDF} & 47.15 & 40.09 & 16.66 & 40.44 & 18.82 & 18.76 & 16.14 & 18.16 \\
ProSDD     & \textit{AffectDF} & 48.06 & 9.60  & 13.87 & 26.92 & 51.86 & 21.89 & 27.47 & 34.18 \\
Voxtral     & \textit{AffectDF} & 6.02 & 0.62 & 1.10 & 3.71 & 34.53  & 4.90 & 25.33 & 24.74 \\
\midrule
Qwen-2.5-Omni & \textit{Inference-only} & 45.51 & 48.29 & 48.90 & 47.57 & 44.64 & 47.17 & 45.60 & 45.87 \\
Qwen-3.0-Omni & \textit{Inference-only} & 36.60 & 53.34 & 45.91 & 46.21 & 26.81 & 43.61 & 50.64 & 40.81 \\
Voxtral        & \textit{Inference-only} & 54.77 & 55.89 & 53.83 & 54.85 & 46.94 & 49.96 & 37.61 & 45.18 \\
\bottomrule
\end{tabular}
\captionsetup{justification=centering}
\label{tab:attackwise_eer}
\end{table*}

\subsection{Emotion-wise Robustness Analysis}
Table~\ref{tab:emotion_results} shows that emotion-wise robustness remains highly inconsistent across datasets, training conditions, and architectures. Models trained on ASVspoof2019 often achieve low EERs on EmoFake, particularly SSL-based systems, but degrade substantially on AffectDF across nearly all emotions. No single emotion is consistently the most difficult; instead, the highest EER shifts across neutral, happiness, anger, sadness, and surprise depending on the model and training data. The training data is balanced across emotional states, as reported in Appendix~\ref{sec:emo_bal}, suggesting that the observed disparities are not simply caused by emotion-level sample imbalance. These results suggest that degradation is driven by the interaction between emotional prosody, attack type, and detector architecture. AffectDF training does not eliminate this variability, showing that current SDD systems still fail to learn emotion-invariant spoof representations. 

\subsection{Performance of LALM-based SDD}

Table~\ref{tab:lalm_results} shows that current inference-only LALM systems exhibit limited robustness for speech deepfake detection. Qwen-2.5-Omni achieves comparatively lower EERs on ASVspoof2019 and EmoFake, but degrades substantially on ASVspoof5 and AffectDF, indicating limited generalization under larger-scale and more diverse emotional spoofing conditions. Qwen-3.0-Omni exhibits inconsistent behavior across datasets, while Voxtral performs poorly across nearly all evaluation conditions. These results suggest that general-purpose LALMs do not detect robust spoof-relevant representations through prompting alone and remain sensitive to attack diversity and emotional variability. The prompts used are provided in Appendix~\ref{sec:appendix_prompts}.

Table~\ref{tab:lalm_results} also reports Voxtral fine-tuning results. When Voxtral is fine-tuned on the conventional ASVspoof2019 benchmark, it performs very well in-domain, but its performance degrades on ASVspoof5 and AffectDF. Its relatively strong performance on EmoFake is consistent with our observation that several models achieve low EERs on EmoFake while still failing on AffectDF, suggesting that robustness on smaller emotional datasets does not necessarily transfer to broader emotional spoofing conditions with higher attack and speaking-style diversity. Conversely, when Voxtral is fine-tuned on AffectDF, it improves over the inference-only setting and also improves on ASVspoof2019, which differs from the overall trend observed for conventional SDD models in Table~\ref{tab:main_results}. However, its performance remains limited on emotional benchmarks and degrades substantially on ASVspoof5. These results show that, even with LALM fine-tuning, spoof-detection performance remains unreliable across paralinguistic and dataset shifts. Additional fine-tuning and inference details are provided in Appendix~\ref{sec:finetuning_details}.

Table~\ref{tab:emotion_results} further provides an emotion-wise view of these LALM-based results. Under inference-only evaluation, Voxtral remains consistently poor across emotional states, while Qwen-2.5-Omni and Qwen-3.0-Omni show more emotion-dependent variation. Fine-tuning Voxtral on ASVspoof2019 improves performance substantially on EmoFake across emotions, but the gains do not transfer consistently to AffectDF. When fine-tuned on AffectDF, Voxtral still shows uneven performance across emotional states on both EmoFake and AffectDF. This is consistent with our finding that, even for LALM-based SDD, no single emotion is uniformly difficult and emotion-wise robustness varies across models and training settings.



\subsection{Acted vs. Spontaneous Emotional Speech}
Table~\ref{tab:attackwise_eer} compares acted and spontaneous emotional speech under matched generation models: A13/A14 use GenVC, A18/A19 use StyleTTS2, and A20/A21 use F5-TTS, while changing the source corpus from acted ESD to spontaneous MSP-Podcast. Overall, the acted--spontaneous gap is highly model- and attack-dependent rather than consistently favoring one speaking style. Models trained on ASVspoof2019 generally struggle more on acted attacks, particularly RawNet2 and AASIST, suggesting that expressive acted prosody is poorly covered by conventional neutral training. In contrast, ASVspoof5 training substantially improves several systems, even for VC-based attacks such as A13 and A14, despite the ASVspoof5 training partition being primarily TTS-based. This behavior is especially visible for XLSR-Mamba and ProSDD, where A13/A14 become substantially easier than the StyleTTS2 and F5-TTS variants. These results suggest that the broader spoofing variability present in ASVspoof5 partially transfers beyond TTS to unseen VC conditions. Training on AffectDF produces a different and more irregular pattern. Although AffectDF training uses acted emotional speech, several models perform better on spontaneous attacks than on their acted counterparts. RawNet2 performs poorly on acted GenVC A13 but achieves very low EERs on A18/A20 and across spontaneous attacks, indicating that robustness does not transfer uniformly across speaking styles or generation systems. XLSR-Mamba similarly degrades on acted GenVC and StyleTTS2 attacks while performing substantially better on spontaneous conditions. In contrast, ProSDD exhibits consistently poor robustness for GenVC under both acted and spontaneous settings. For LALM-based SDD, fine-tuned Voxtral also shows an acted--spontaneous gap. Fine-tuning on ASVspoof2019 yields lower EERs on acted than spontaneous attacks, and this trend largely remains with AffectDF fine-tuning. 

These attack-level inconsistencies suggest that emotional training alone does not guarantee style-invariant spoof detection. Robustness depends strongly on the interaction among source speech style, generation model, and detector architecture.
\vspace{-2mm}
\paragraph{Summary of Research Questions.}
The findings can be mapped to the research questions in Section~1. For RQ1, models trained on conventional benchmarks show substantial degradation on AffectDF despite strong performance on conventional and smaller emotional benchmarks, demonstrating limited robustness to emotional variability, attack diversity, and speaking-style shifts. For RQ2, training on AffectDF does not consistently improve cross-domain generalization and often trades gains on emotional data for degraded performance on conventional benchmarks, indicating that emotional training alone is insufficient. For RQ3, performance varies across emotional states, attack families, and acted versus spontaneous speech, with the effect depending on the model architecture and training condition. These findings establish AffectDF as a challenging benchmark for developing modern training strategies that learn more discriminative and transferable spoof cues across both conventional and emotional spoofing conditions.

\section{Conclusion}
We introduced AffectDF, a comprehensive benchmark for emotionally expressive speech deepfake detection, spanning acted and spontaneous emotional speech across TTS, VC, EVC, and LALM-based attacks. Our experiments show that current SDD systems trained on conventional benchmarks degrade substantially under emotional spoofing conditions, with several systems approaching near-random performance on AffectDF. While emotional training improves robustness in some cases, the gains remain inconsistent across architectures and often reduce generalization to conventional benchmarks. We further show that robustness varies across emotional states, attack families, and speaking styles, indicating that current SDD systems still struggle to learn generalized spoof-relevant representations under emotionally expressive conditions. Overall, AffectDF exposes important limitations of existing SDD systems and provides a benchmark for developing more robust and transferable speech deepfake detectors.

\section*{Limitations}

While this work demonstrates the importance of AffectDF and highlights the vulnerability of current SDD systems to emotionally expressive synthetic speech across multiple architectures, training settings, and evaluation conditions, some additional experiments could further strengthen the analysis. First, the speaker diversity in the AffectDF training set is limited. This is partly due to the small number of English speakers in ESD and our need to maintain disjoint acted and spontaneous speech conditions across the train and test partitions. Second, although we evaluate both inference-only LALMs and fine-tuned Voxtral, our fine-tuning study is limited to a single LALM. Fine-tuning a broader set of LALMs across different prompts and adaptation strategies could provide a more complete understanding of their potential for SDD (Additional prompt and fine-tuning details are provided in Appendices~\ref{sec:appendix_prompts} and~\ref{sec:finetuning_details}). Third, we do not conduct an extensive prompt-sensitivity analysis for LALM-based detection. Future work could explore emotion-aware prompts, attack-aware prompts, and multi-prompt calibration strategies to better understand how prompting affects LALM-based spoof detection under emotionally expressive conditions. Finally, although our acted versus spontaneous analysis focuses on the broader distinction between controlled acted emotional speech and naturalistic spontaneous emotional speech, ESD and MSP-Podcast also differ in other corpus-level factors, as detailed in Appendix \ref{sec:MSPvsESD}. Future work will investigate these factors more explicitly to better isolate their individual effects on emotional speech deepfake detection.

\section*{Ethical Considerations}
In this work, we use publicly available datasets and models to generate synthetic speech for the purpose of improving speech deepfake detection. We do not release any additional speaker information beyond what is already available in the original source datasets, and we do not synthesize samples containing harmful or sensitive content. The generated data and analyses are intended solely to support research on robust detection, security, and responsible benchmarking of speech deepfakes. We do not intend for AffectDF or any generation pipeline described in this work to be used for impersonation, deception, or other malicious applications.


\section{Acknowledgments}
This work was supported by 
\begin{itemize}
    \item The National Science Foundation (NSF) CAREER Award IIS-2533652, and 
    \item The Office of the Director of National Intelligence (ODNI), Intelligence Advanced Research Projects Activity (IARPA), via the ARTS Program, Contract \#D2023-2308110001
\end{itemize}
The views and conclusions contained herein are those of the authors and should not be interpreted as necessarily representing the official policies, either expressed or implied, of ODNI, IARPA, or the U.S. Government. The U.S. Government is authorized to reproduce and distribute reprints for governmental purposes notwithstanding any copyright annotation therein.

\bibliography{mybib}
\newpage
\begin{table*}[t]
\centering
\renewcommand{\arraystretch}{1.2}
\setlength{\tabcolsep}{5pt}
\small
\caption{Comparison of AffectDF with existing conventional and emotional speech deepfake detection benchmarks. ``Seen'' indicates that the development set uses the same attack systems as the training set.}
\begin{tabularx}{\textwidth}{lccc cccc c X}
\hline
\multirow{2}{*}{\textbf{Dataset}} 
& \multicolumn{3}{c}{\textbf{No. of Samples}} 
& \multicolumn{4}{c}{\textbf{No. of Attacks}} 
& \multirow{2}{*}{\textbf{Hours ($\sim$)}} 
& \multirow{2}{*}{\textbf{Attack Types}} \\
\cline{2-8}
& \textbf{Train} & \textbf{Dev} & \textbf{Test}
& \textbf{Train} & \textbf{Dev} & \textbf{Test} & \textbf{Total}
&  &  \\
\hline
ASVspoof2019-LA 
& 25,380 & 24,844 & 71,237
& 6 & Seen & 13 & 19
& 111 
& TTS, VC \\

ASVspoof5-Track1 
& 182,358 & 140,950 & 680,774
& 8 & 8 & 16 & 32
& 2000 
& TTS, VC, Adversarial \\

EmoFake 
& 27,300 & 9,100 & 17,500
& 2 & Seen & 5 & 7
& 42  
& EVC \\

EmoSpoof-TTS 
& 9,600 & 4,800 & 9,600
& 1 & 1 & 1 & 3
& 29 
& TTS \\

AffectDF 
& 86,999 & 23,330 & 175,468
& 5 & 2 & 14 & 21
& 260 
& TTS, VC, EVC, LALM\\
\hline
\end{tabularx}
\label{tab:dataset_comparison}
\end{table*}
\appendix

\section{AffectDF-Details}
\label{sec:affectDF_deets}
\subsection{Data Split Configurations}
\label{sec:data_split}
AffectDF is the largest and most diverse emotional speech deepfake dataset for SDD. Table~\ref{tab:dataset_comparison} compares AffectDF with existing conventional benchmarks and emotional spoofing datasets in terms of sample size, attack diversity, duration, and attack types. Compared with widely used conventional benchmarks such as ASVspoof2019-LA, AffectDF contains substantially more samples and a broader range of spoofing attacks. Existing emotional spoofing datasets, including EmoFake and EmoSpoof-TTS, are considerably smaller in both scale and attack diversity, typically focusing on a single generation paradigm such as EVC or TTS. In contrast, AffectDF spans TTS, VC, EVC and LALM-EVC attacks, covering 21 attack systems across five emotional states and both acted and spontaneous emotional speech. Although ASVspoof5 remains larger in total sample count, AffectDF provides comparable attack-scale diversity while introducing emotionally expressive and LALM-based attacks that are absent from prior SDD benchmarks. These properties make AffectDF a comprehensive benchmark for evaluating robustness to modern emotionally expressive spoofing conditions.

\subsection{Attacks Details}
\label{sec:attack_deets}
This section provides additional details on the generation setup used for AffectDF. For TTS, VC, and EVC attacks, a subset of samples from the main dataset was held out exclusively as reference audio to prevent overlap with the utterances used for generation. Links to the implementation pages/models is listed in Table \ref{table:models}

\subsubsection{LALM-EVC models}
\paragraph{Qwen2.5-Omni.} \cite{qwen2.5omni}
Qwen2.5-Omni-7B is an end-to-end omni-modal model based on a Thinker--Talker architecture. It uses Time-aligned Multimodal RoPE to jointly process text, vision, and audio streams, and can autoregressively generate synchronized text and real-time speech using a sliding-window diffusion transformer.
\paragraph{Kimi-Audio.} \cite{kimi}
Kimi-Audio is pre-trained on more than 13 million hours of diverse audio and text data. It uses a hybrid continuous--discrete acoustic representation, together with a core language model, parallel generation heads, and a flow-matching streaming detokenizer for unified and low-latency audio understanding and generation.
\begin{table}[ht]
\centering
\caption{
Sources of the employed generation models.
}
\resizebox{\columnwidth}{!}{%
\begin{tabular}{@{}ll@{}}
\toprule
\textbf{Models}          & \textbf{Sources}            \\ \midrule
Qwen2.5-Omni-7B          & \url{https://huggingface.co/Qwen/Qwen2.5-Omni-7B}               \\
Kimi-Audio               & \url{https://huggingface.co/moonshotai/Kimi-Audio-7B-Instruct}  \\
MiniCPM-o 4.5            & \url{https://huggingface.co/openbmb/MiniCPM-o-4_5}              \\
F5-TTS               & \url{https://github.com/swivid/f5-tts}  \\
StyleTTS2              & \url{https://github.com/yl4579/StyleTTS2}  \\
CosyVoice              & \url{https://github.com/FunAudioLLM/CosyVoice}  \\
CosyVoice2                & \url{https://huggingface.co/FunAudioLLM/CosyVoice2-0.5B}  \\
CosyVoice3             & \url{https://huggingface.co/FunAudioLLM/Fun-CosyVoice3-0.5B-2512}  \\
Qwen3-TTS              & \url{https://huggingface.co/Qwen/Qwen3-TTS-12Hz-1.7B-Base}  \\
Vevo2             & \url{https://huggingface.co/RMSnow/Vevo2}  \\
GenVC             & \url{https://github.com/caizexin/GenVC}  \\
TriAAN-VC                & \url{https://github.com/winddori2002/TriAAN-VC}  \\
DDDMVC             & \url{https://github.com/hayeong0/DDDM-VC}  \\
\bottomrule
\end{tabular}%
}

\label{table:models}
\end{table}

\paragraph{MiniCPM-o 4.5.} \cite{minicpm}
MiniCPM-o 4.5 is a 9B-parameter full-duplex omni-modal model designed for real-time proactive interaction. It adopts an Omni-Flow time-division multiplexing mechanism to align multimodal streams on a shared timeline, and integrates a Qwen3-8B backbone with SigLIP2 and Whisper-medium encoders through continuous token-level hidden connections.

\paragraph{Steered LALM-EVC generation.} In addition to standard instruction-following LALM-EVC generation, AffectDF includes neuron-steered LALM-EVC attacks for Qwen2.5-Omni and Kimi-Audio. These samples are generated using the emotion-sensitive neuron (ESN) intervention framework~\citep{zhao2026discoveringcausallyvalidatingemotionsensitive, zhao2026neuronlevelemotioncontrolspeechgenerative}. Specifically, ESN masks are identified from successful emotional voice conversion instances that satisfy both target-emotion realization and linguistic content preservation, and are then applied at inference time through activation steering without updating model parameters. We use the same EVC prompt as the standard LALM-EVC setting  (prompts used can be found in Appendix~\ref{sec:appendix_prompts}), while activating the ESN mask corresponding to the target emotion during generation. This produces emotionally steered LALM-based spoofing attacks, denoted as Qwen2.5-Omni (steered) and Kimi-Audio (steered) in Table~\ref{tab:dataset_split}, enabling AffectDF to evaluate SDD robustness not only against prompt-controlled LALM generation but also against more explicitly emotion-controlled LALM-based attacks.



\subsubsection{TTS-Models}
\paragraph{F5-TTS.}\cite{F5TTS}
F5-TTS is a fully non-autoregressive text-to-speech system based on flow matching with a Diffusion Transformer (DiT). Unlike conventional TTS pipelines that rely on explicit duration modeling, text encoders, or phoneme-level alignment, F5-TTS pads the text input to match the speech length and performs denoising-based speech generation. In our experiments, we use the official F5-TTS inference pipeline with reference audio and target text to generate emotionally expressive synthetic speech.
\paragraph{Style-TTS2.}\cite{StyleTTS2}
StyleTTS2 is a text-to-speech model that improves naturalness and speaker adaptation through style diffusion and adversarial training with large speech language model discriminators. It models speaking style as a latent variable and uses diffusion-based sampling to generate suitable style representations for the target text, enabling expressive and natural speech synthesis. In our experiments, we use the official StyleTTS2 inference pipeline with reference speech and target text to generate emotional synthetic speech.
\paragraph{CosyVoice.} \cite{Cosyvoice}
CosyVoice is a scalable multilingual zero-shot text-to-speech system based on supervised semantic speech tokens. It models speech generation in two stages: an LLM generates speech tokens conditioned on the input text and reference speech, and a conditional flow-matching model synthesizes the waveform from the generated tokens. In our experiments, we use the official CosyVoice zero-shot inference pipeline with reference audio and target text to generate emotionally expressive synthetic speech.
\paragraph{CosyVoice2.} \cite{cosyvoice2}
CosyVoice 2 is a scalable multilingual streaming text-to-speech system built on supervised discrete speech tokens with large language model backbones. It extends the two-stage generation framework of its predecessor by introducing finite-scalar quantization to improve codebook utilization, and streamlines the text-speech LM to directly leverage a pre-trained LLM as the backbone. Waveform synthesis is performed by a chunk-aware causal flow matching model, which supports both streaming and non-streaming inference within a single unified model. In our experiments, we use the official CosyVoice 2 zero-shot inference pipeline with reference audio and target text to generate emotionally expressive synthetic speech.
\paragraph{CosyVoice3.} \cite{cosyvoice3}
CosyVoice 3 is zero-shot multilingual speech synthesis model designed for in-the-wild generation, surpassing its predecessor in content consistency, speaker similarity, and prosody naturalness. It retains the two-stage LLM and flow matching framework of CosyVoice 2, while introducing several key advances: a novel speech tokenizer for improved prosody, developed via supervised multi-task training across ASR, speech emotion recognition, language identification, audio event detection, and speaker analysis; a differentiable reward model for post-training; and a substantial scaling of training data from ten thousand to one million hours, covering 9 languages and 18 Chinese dialects. In our experiments, we use the official CosyVoice 3 zero-shot inference pipeline with reference audio and target text to generate emotionally expressive synthetic speech.
\paragraph{Qwen3-TTS.} \cite{qwen3tts}
Qwen3-TTS is a multilingual, controllable, streaming text-to-speech system built on the codec language modeling paradigm, supporting natural language–based style control and voice cloning from short reference audio. In this work, we leverage its high-fidelity voice cloning mode using emotionally expressive reference speech paired with target text for synthesis. The system employs the Qwen-TTS-Tokenizer-12.5Hz — a 16-layer multi-codebook speech tokenizer with a lightweight causal ConvNet for encoding and decoding — paired with a 1.7B-parameter language model. A Multi-Token Prediction (MTP) module jointly models multi-codebook speech tokens and enables immediate decoding from the first codec frame, supporting low-latency streaming generation. Trained on over 5 million hours of multilingual speech data, Qwen3-TTS achieves robust, high-quality synthesis across diverse speakers and languages.

\subsubsection{VC and EVC-Models}
\paragraph{Vevo2.} \cite{vevo2} Vevo2 is a unified speech and singing generation model that supports multiple tasks, including zero-shot TTS, singing voice synthesis, voice conversion, speech/singing editing, and melody-conditioned generation. The released models are trained on large-scale speech and singing corpora, including approximately 101k hours of multilingual speech from Emilia~\cite{emilia} and around 7k hours of singing voice data from SingNet-7k~\cite{singnet}, followed by post-training on INTP~\cite{INTP} and M4Singer~\cite{m4singer}. Architecturally, Vevo2 performs joint prosody modeling using two discrete tokenizers: a prosody tokenizer for melody and rhythm, and a content-style tokenizer for linguistic content and speaking/singing style while disentangling speaker timbre. Generation is performed with an autoregressive content-style language model based on Qwen2.5-0.5B~\cite{qwen2.5}, followed by a flow-matching acoustic model and a Vocos-based neural vocoder for waveform synthesis.

\paragraph{GenVC.} \cite{genvc} 
We use GenVC for both VC and EVC attacks. GenVC is a self-supervised zero-shot voice conversion model based on discrete speech token modeling. It extracts content and acoustic tokens from speech, encodes speaker and style information from a reference utterance using a Perceiver encoder, and generates target acoustic tokens with a GPT-style autoregressive decoder, which are then converted to waveform using a HiFiGAN vocoder. In our experiments, we use the GenVC-Large checkpoint, which is initialized from GenVC-Small and further trained with additional English speech from CommonVoice-EN and MLS-EN.

\paragraph{ConsistencyVC.} \cite{consistencyVC} ConsistencyVC is a VITS-style voice conversion model that combines a content encoder, a jointly trained speaker encoder, a flow module, a posterior encoder, and a HiFi-GAN-based decoder. The model uses PPG features for expressive voice conversion, while the speaker encoder extracts reference-speaker information from the mel-spectrogram. Its key idea is to apply a speaker consistency loss directly to the jointly trained speaker encoder, improving speaker similarity while helping preserve emotional characteristics from the reference speech. In our experiments, we use a ConsistencyVC model trained on the full NaturalVoices subset.

\paragraph{TriAAN-VC.} \cite{consistencyVC} TriAAN-VC is an any-to-any voice conversion model composed of a content encoder, a speaker encoder, and a decoder with a Triple Adaptive Attention Normalization module. The content encoder extracts linguistic content from the source speech, while the speaker encoder captures speaker information from the target speech. The TriAAN block injects target-speaker information into the source content through channel-wise, temporal, and global/statistical adaptive normalization, while a siamese loss helps preserve source content and balance intelligibility with speaker similarity. In our experiments, we use a TriAAN-VC model trained on the full NaturalVoices subset.

\paragraph{DDDMVC.} \cite{dddmvc} DDDM-VC is a diffusion-based voice conversion model that uses disentangled speech representations to separately control linguistic content, intonation, and timbre. It first extracts self-supervised speech representations and then applies decoupled denoising diffusion models to resynthesize speech while denoising different speech attributes separately. The model also introduces prior mixup, where a mixed-style converted representation is used as the diffusion prior to improve robust voice-style transfer. In our experiments, we use a DDDM-VC model trained on the 50\% NaturalVoices subset.

\begin{table}[t]
\centering
\small
\caption{Emotion-wise number of samples in AffectDF splits. Counts include both fake and real samples.}
\label{tab:affectdf-split-emotion-counts}
\resizebox{\columnwidth}{!}{
\begin{tabular}{lrrrrrr}
\toprule
Split & Neutral & Happy & Angry & Sad & Surprise & Total \\
\midrule
Train & 11800 & 18800 & 18800 & 18800 & 18799 & 86999 \\
Dev   & 4663  & 4667  & 4681  & 4653  & 4666  & 23330 \\
Test  & 34536 & 38541 & 34374 & 33865 & 34152 & 175468 \\
\midrule
Total & 50999 & 62008 & 57855 & 57318 & 57617 & 285797 \\
\bottomrule
\end{tabular}
}
\end{table}

\begin{table}[t]
\centering
\small
\caption{Dataset statistics across training, development, and test splits.}
\label{tab:dataset-statistics}
\resizebox{\columnwidth}{!}{
\begin{tabular}{llrrrr}
\toprule
Dataset & Split & Spoof & Real & Total & Spoof:Real \\
\midrule
\multirow{3}{*}{AffectDF}
& Train & 80,999  & 6,000   & 86,999  & 93:07 \\
& Dev   & 20,330  & 3,000   & 23,330  & 87:13 \\
& Test  & 165,360 & 10,108  & 175,468 & 94:06 \\
\midrule
\multirow{3}{*}{ASVspoof2019}
& Train & 22,800  & 2,580   & 25,380  & 90:10 \\
& Dev   & 22,296  & 2,548   & 24,844  & 90:10 \\
& Test  & 63,882  & 7,355   & 71,237  & 90:10 \\
\midrule
\multirow{3}{*}{ASVspoof5}
& Train & 163,560 & 18,797  & 182,357 & 90:10 \\
& Dev   & 109,616 & 31,334  & 140,950 & 78:22 \\
& Test  & 542,086 & 138,688 & 680,774 & 80:20 \\
\bottomrule
\end{tabular}
}
\end{table}

\subsubsection{Emotion Sample Balance}
\label{sec:emo_bal}
As shown in Section~5, SDD performance varies substantially across emotional states. However, Table~\ref{tab:affectdf-split-emotion-counts} shows that AffectDF is balanced across emotion categories within each split. Therefore, the observed emotion-wise performance disparities cannot be attributed simply to sample imbalance. Instead, they suggest that different emotional states introduce distinct acoustic, prosodic, and paralinguistic variations that affect spoof detection in different ways.

\subsubsection{Spoof-to-Real Ratio}
\label{sec:spoof_real}
Table~\ref{tab:dataset-statistics} reports the spoof-to-real ratio of AffectDF and existing SDD benchmarks. We design AffectDF to follow a similar protocol to conventional benchmarks, particularly in the training split, where spoofed samples substantially outnumber bona fide samples. This design helps AffectDF remain compatible with common SDD evaluation settings while enabling controlled analysis under emotionally expressive spoofing conditions.

\section{Baselines and Dataset}
\subsection{SDD Datasets}
\label{sec:appendix_datasets}

\begin{itemize}
   \item \textbf{ASVspoof2019} \cite{ASVspoof2019}: A widely used anti-spoofing benchmark containing attacks from 19 TTS and VC systems. We use the official train, development, and evaluation partitions.

    \item \textbf{ASVspoof2021} \cite{ASVspoof2021}: An evaluation-only benchmark derived from ASVspoof2019, where audio samples are transmitted through real telephony systems, including VoIP and PSTN, to introduce channel variability.

   \item \textbf{ASVspoof5} \cite{ASVspoof2024}: The most recent ASVspoof challenge dataset, incorporating 32 spoofing systems, including TTS, VC, and adversarial attacks, under diverse codec and compression conditions. We use the Track~1 train, development, and evaluation partitions.

    \item \textbf{EmoFake} \cite{EmoFake}: An emotional spoofing dataset constructed using seven emotional VC systems, with bona fide samples drawn from the Emotional Speech Database (ESD) \cite{ESD}. We use the English evaluation partition.

    \item \textbf{EmoSpoof-TTS} \cite{emotionAS}: An emotional spoofing dataset constructed using three TTS systems, with bona fide samples drawn from ESD \cite{ESD}. We use the test partition, which has one unseen attack.
\end{itemize}
\begin{table}[b]
\centering
\caption{
Sources of the employed detection models.
}
\resizebox{\columnwidth}{!}{%
\begin{tabular}{@{}ll@{}}
\toprule
\textbf{Models}          & \textbf{Sources}            \\ \midrule
Qwen3-Omni-30B         & \url{https://huggingface.co/Qwen/Qwen3-Omni-30B-A3B-Instruct}               \\
Qwen2.5-Omni-7B               & \url{https://huggingface.co/Qwen/Qwen2.5-Omni-7B}  \\
Voxtral-Mini-3B           & \url{https://huggingface.co/mistralai/Voxtral-Mini-3B-2507}              \\
RawNet2             & \url{https://github.com/asvspoof-challenge/2021/tree/main/LA/Baseline-RawNet2}  \\
AASIST              & \url{https://github.com/clovaai/aasist}  \\
XLSR-SLS              & \url{https://github.com/QiShanZhang/SLSforASVspoof-2021-DF}  \\
XLSR-MAMBA               & \url{https://github.com/swagshaw/XLSR-Mamba}  \\
ProSDD            & \url{https://prosdd.github.io/ProSDD_website/}  \\
\bottomrule
\end{tabular}%
}
\label{table:models_det}
\end{table}

\subsection{SDD Baseline Models}
\label{sec:appendix_baselines}

\begin{itemize}

\item \textbf{RawNet2} \cite{RawNet2}: An end-to-end anti-spoofing model that operates directly on raw waveforms and has shown strong performance across conventional SDD benchmarks. We use the official pre-trained ASVspoof2019 model and follow the original training protocol for ASVspoof5 and AffectDF.

\item \textbf{AASIST} \cite{AASIST}: An end-to-end SDD architecture based on integrated spectro-temporal graph attention networks for modeling spectral and temporal spoofing artifacts. We use the official ASVspoof2019 pre-trained model and follow the original training setup for ASVspoof5 and AffectDF.

\item \textbf{XLSR-SLS} \cite{ssl_sls}: An SSL-based SDD system that combines XLS-R representations with a Selective Layer Selection (SLS) module to leverage information from multiple transformer layers. The model has shown strong generalization across ASVspoof2019, ASVspoof2021, and in-the-wild benchmarks. We follow the original training setup for ASVspoof2019, ASVspoof5, and AffectDF.

\item \textbf{XLSR-Mamba} \cite{xlsr-mamba}: An SSL-based SDD system that combines self-supervised speech representations with a dual-column Mamba architecture for spoof detection. The model has shown strong generalization across ASVspoof2021 and in-the-wild benchmarks when trained on ASVspoof2019. We follow the original training setup for ASVspoof2019, ASVspoof5, and AffectDF.

\item \textbf{Prompt-based LALM Evaluation} \cite{qwen2.5omni, qwen3omni, liu2025voxtral}: We evaluate recent LALMs using prompt-based inference for SDD. The same prompt is used across all models, where the input audio is provided to the model and spoof probabilities between 0 and 1 are generated following the prompt instructions.

\item \textbf{Fine-tuned LALM for SDD}  \cite{liu2025voxtral}: We further fine-tune Voxtral for SDD using supervised LoRA with bona fide/spoof labels and cross-entropy loss. During inference, instead of generating a floating-point score, we compute the negative log-likelihoods of the candidate tokens `0' and `1' and apply a two-class softmax to obtain the spoof probability. Additional fine-tuning and inference details are provided in Appendix~\ref{sec:finetuning_details}.
\end{itemize}

Links to the implementation pages/models
is listed in Table \ref{table:models_det}. The prompt used for LALM-based SDD is listed in Appendix \ref{sec:appendix_prompts}

\subsection{Model Training and Inference Details}
\label{sec:model_training_details}

To improve reproducibility, we provide the model-specific audio preprocessing, training, checkpoint selection, and inference settings used for each evaluated SDD system. For trainable baselines, we follow the original training configurations while applying the same label-independent preprocessing to real and fake samples within each model pipeline. For inference-only LALM systems, we report the prompt-based scoring setup. For Voxtral-FT, we additionally summarize the supervised LoRA fine-tuning and NLL-based inference strategy, with further details provided in Appendix~\ref{sec:finetuning_details}. Table~\ref{tab:reproducibility_details} summarizes these settings.

\begin{table*}[t]
\centering
\small
\caption{
Audio preprocessing, training, and inference details for SDD systems.
}
\setlength{\tabcolsep}{5pt}
\renewcommand{\arraystretch}{1.2}
\begin{tabular}{p{0.16\textwidth}p{0.76\textwidth}}
\toprule
\textbf{Model} & \textbf{Reproducibility Details} \\
\midrule

RawNet2 &
\textbf{Audio:} Loaded as 16~kHz mono and padded/cropped to 64,600 samples ($\sim$4~s).
\textbf{Training:} 50 epochs; batch size 32; Adam optimizer; learning rate $1\times10^{-4}$; weight decay $1\times10^{-4}$; weighted cross-entropy; best checkpoint selected based on the lowest development loss. \\

\midrule
AASIST &
\textbf{Audio:} Converted to mono, resampled to 16~kHz, and padded/cropped to 64,600 samples.
\textbf{Training:} 50 epochs; batch size 32; Adam optimizer; learning rate $1\times10^{-4}$; weight decay $1\times10^{-4}$; weighted cross-entropy; best checkpoint selected based on the lowest development loss. \\

\midrule
XLSR-SLS &
\textbf{Audio:} Loaded as 16~kHz mono and padded/cropped to 64,600 samples ($\sim$4~s). RawBoost augmentation is used during training.
\textbf{Training:} Up to 50 epochs; batch size 5; learning rate $1\times10^{-6}$; weight decay $1\times10^{-4}$; weighted cross-entropy. \\

\midrule
XLSR-Mamba &
\textbf{Audio:} Loaded as 16~kHz mono and padded/cropped to 66,800 samples ($\sim$4.2~s). RawBoost augmentation is used during training.
\textbf{Training:} 7 epochs; batch size 20; learning rate $1\times10^{-6}$; weight decay $1\times10^{-4}$; weighted cross-entropy; final-epoch checkpoint used for evaluation. \\

\midrule
ProSDD &
\textbf{Audio:} Converted to mono, resampled to 16~kHz, and padded/cropped to 64,000 samples (4~s). RawBoost is applied to 50\% of samples during Stage-II training.
\textbf{Training:} Initialized from the released Stage-I checkpoint and trained for Stage II for 50 epochs; batch size 64; learning rates $1\times10^{-6}$, $1\times10^{-4}$, and $1\times10^{-5}$ for the XLS-R backbone, projection layer, and classifier, respectively; weight decay $1\times10^{-4}$; weighted cross-entropy and supervised masked-prediction losses. Speaker and prosodic embeddings are extracted using the released code. \\

\midrule
Qwen-2.5-Omni &
The input audio is passed to the model processor. At inference, the model is prompted to generate a single score in $[0,1]$, interpreted as the probability of fake, where 0 denotes real and 1 denotes fake. A maximum of 6 generated tokens is used, and three inference runs are conducted. \\

\midrule
Qwen-3.0-Omni &
The input audio is passed to the model processor. At inference, the model is prompted to generate a single score in $[0,1]$, interpreted as the probability of fake, where 0 denotes real and 1 denotes fake. Sampled decoding is used with temperature 0.7, top-$p=0.9$, and a maximum of 6 generated tokens; three inference runs are conducted. \\

\midrule
Voxtral &
\textbf{Audio:} Loaded as mono 24~kHz audio and passed through the model's audio encoder.
\textbf{Inference:} The model is prompted to generate a single score in $[0,1]$, interpreted as the probability of fake, where 0 denotes real and 1 denotes fake. Sampled decoding is used with temperature 0.2, top-$p=0.95$, and a maximum of 10 generated tokens; three inference runs are conducted. \\

\midrule
Voxtral-FT &
\textbf{Audio:} Loaded as mono 24~kHz audio and passed through the model's audio encoder.
\textbf{Training:} Supervised LoRA fine-tuning for 3 epochs with token-level cross-entropy; batch size 1/GPU with 16-step gradient accumulation; learning rate $1\times10^{-5}$; cosine scheduling with 3\% warm-up; LoRA rank $r=16$, $\alpha=32$, and dropout 0.05.
\textbf{Inference:} NLLs of candidate outputs `0' and `1' are computed from token logits, followed by a two-class softmax over negative NLLs to obtain the spoof probability. See Appendix~\ref{sec:finetuning_details}. \\

\bottomrule
\end{tabular}
\label{tab:reproducibility_details}
\end{table*}

\section{Prompts}
\label{sec:appendix_prompts}
\subsection{Generation}
\label{sec:appendix_prompts_gen}
\begin{lstlisting}[style=promptbox, caption={Prompt used for EVC generation of LALMs.}, label={lst:evc-prompt}]
Repeat the audio word-for-word exactly.
Speak with {target_emotion} emotion.
Do not add or remove any words. Output speech only.
\end{lstlisting}

\subsection{Detection}
\label{sec:appendix_prompts_detect}

\begin{lstlisting}[
style=promptbox,
caption={Prompt used for LALM-based speech deepfake detection.},
label={lst:evc-prompt},
breaklines=true,
breakatwhitespace=true,
breakindent=0pt,
breakautoindent=false,
postbreak={},
columns=fullflexible,
keepspaces=true
]
You are an audio deepfake detector.
Listen to the input audio and estimate the probability that it is fake or real.
Return only one floating point number between 0 and 1.
0 means definitely real.
1 means definitely fake.
Do not output any explanation, words, or extra symbols.
\end{lstlisting}













\begin{table}[t]
\centering
\scriptsize
\caption{
EmoSpoof-TTS evaluation across training conditions.
Results are reported in EER (\%).
Neutral (N), Happy (H), Angry (A), Sad (S), Surprise (Su).
}
\setlength{\tabcolsep}{3.5pt}
\renewcommand{\arraystretch}{1.2}
\begin{tabular}{llccccc}
\toprule
Model & Train Data & O & N & H & A & S \\
\midrule
RawNet2    & \textit{ASVspoof2019} & 43.04 & 39.00 & 45.00 & 46.75 & 35.17 \\
AASIST     & \textit{ASVspoof2019} & 31.06 & 27.17 & 35.67 & 32.58 & 25.42 \\
XLSR-SLS   & \textit{ASVspoof2019} & 18.92 & 15.67 & 24.33 & 15.92 & 15.92 \\
XLSR-M     & \textit{ASVspoof2019} & 14.31 & 13.25 & 18.33 & 12.58 & 10.58 \\
ProSDD     & \textit{ASVspoof2019} & 9.54  & 6.92  & 12.92 & 9.50  & 7.75 \\
\midrule
RawNet2    & \textit{ASVspoof5} & 27.13 & 25.42 & 27.08 & 26.29 & 25.33 \\
AASIST     & \textit{ASVspoof5} & 15.19 & 14.92 & 13.67 & 8.58  & 18.58 \\
XLSR-SLS   & \textit{ASVspoof5} & 25.92 & 27.25 & 22.92 & 21.75 & 29.00 \\
XLSR-M     & \textit{ASVspoof5} & 17.40 & 15.17 & 20.42 & 13.33 & 18.92 \\
ProSDD     & \textit{ASVspoof5} & 11.96 & 11.17 & 11.08 & 8.67  & 13.33 \\
\midrule
RawNet2    & \textit{AffectDF} & 0.44  & 0.00  & 0.67  & 0.17  & 0.08 \\
AASIST     & \textit{AffectDF} & 4.83  & 5.42  & 4.17  & 2.92  & 6.83 \\
XLSR-SLS   & \textit{AffectDF} & 3.38  & 2.83  & 3.33  & 2.42  & 4.92 \\
XLSR-M     & \textit{AffectDF} & 3.35  & 1.92  & 2.92  & 2.58  & 4.67 \\
ProSDD     & \textit{AffectDF} & 5.46  & 2.42  & 2.42  & 2.58  & 13.92 \\
\midrule
Qwen-2.5-Omni & \textit{Inference-only} & 20.51 & 22.80 & 22.84 & 20.24 & 14.05 \\
Qwen-3.0-Omni & \textit{Inference-only} & 55.21 & 55.31 & 54.22 & 54.63 & 57.18 \\
Voxtral        & \textit{Inference-only} & 51.44 & 50.76 & 51.61 & 52.31 & 51.00 \\
\bottomrule
\end{tabular}
\label{tab:emospoof_results}
\end{table}

\section{LALM Fine-tuning for SDD}
\label{sec:finetuning_details}
We fine-tune Voxtral-Mini-3B for speech deepfake detection using parameter-efficient LoRA adaptation. For each training sample, the model receives the audio input together with a fixed  prompt as mentioned in Appendix \ref{sec:appendix_prompts_detect} and is trained to output a single target token, where `0' denotes real and `1' denotes fake. Audio is loaded as mono 24~kHz input and passed through Voxtral's audio encoder. We optimize only the LoRA parameters while keeping the base Voxtral model frozen. Following prior LALM fine-tuning settings for SDD~\cite{LLM-SDD}, we train for 3 epochs with a learning rate of $1\times10^{-5}$, cosine scheduling with 3\% warm-up, batch size 1 per GPU, 16-step gradient accumulation, and AdamW optimization. We use LoRA rank $r=16$, scaling factor $\alpha=32$, and dropout 0.05. LoRA adapters are applied to the matching projection modules in Voxtral, including the attention projections and feed-forward projections. The best checkpoint is selected using the lowest development loss.

During training, the loss is computed only over the assistant response token and not over the prompt or audio-context tokens. Specifically, prompt tokens are masked from the label sequence, and token-level cross-entropy is applied only to the target output token `0' or `1'. This directly trains the model for binary spoof detection while preserving the original instruction-following format of the LALM.

At inference time, we avoid asking the model to generate a continuous spoof probability, since LALMs can be unreliable at producing calibrated floating-point values. Instead, for each test audio, we score the two candidate outputs `0' and `1' by computing their negative log-likelihoods from the model logits. We then apply a two-class softmax over the negative NLLs and use the normalized probability assigned to `1' as the spoof score. This score is used to compute EER for all evaluation datasets.

\section{Impact of Emotion on SDD: EmoSpoof-TTS}
\label{sec:emospooftts}

We repeat the main evaluation setup on EmoSpoof-TTS and report the results in Table~\ref{tab:emospoof_results}. Since the EmoSpoof-TTS test partition contains only one unseen TTS attack, this analysis is less comprehensive than AffectDF, but it shows consistent trends. Under conventional training, RawNet2 and AASIST degrade substantially, while ProSDD achieves the strongest performance among ASVspoof-trained systems, with EERs of 9.54\% and 11.96\% when trained on ASVspoof2019 and ASVspoof5, respectively. In contrast, training on AffectDF yields much lower EERs across most models, with RawNet2 achieving the best overall result of 0.44\%. This suggests that AffectDF provides useful transferable cues for emotional TTS detection, even though the gain is still architecture-dependent. Emotion-wise performance also varies across models and training settings, reinforcing our main finding that emotional spoof detection is not governed by a single difficult emotion, but by the interaction between training data, attack type, and detector architecture.

\section{Additional Evaluation Metric}
\label{sec:tDCF}

We additionally report minimum tandem detection cost function (min t-DCF) on the ASVspoof2019 evaluation set to complement the EER-based analysis. Since min t-DCF requires ASV system scores, we compute it only for ASVspoof2019, where the official ASV scores and protocol are available. As shown in Table~\ref{tab:tdcf_results}, the min t-DCF results are consistent with the EER trends: models with low EER also obtain low min t-DCF, while models with poor cross-dataset generalization obtain min t-DCF values close to 1.

\begin{table}[t]
\centering
\small
\caption{
ASVspoof2019 evaluation using EER and min t-DCF.
}
\setlength{\tabcolsep}{4pt}
\renewcommand{\arraystretch}{1.1}
\begin{tabular}{llcc}
\toprule
Model & Train & EER & min t-DCF \\
\midrule
RawNet2 & ASV19 & 4.60 & 0.11 \\
RawNet2 & ASV5 & 24.75 & 0.51 \\
RawNet2 & AffectDF & 43.02 & 0.94 \\
\midrule
AASIST & ASV19 & 0.83 & 0.027 \\
AASIST & ASV5 & 23.16 & 0.44 \\
AASIST & AffectDF & 44.52 & 1.00 \\
\midrule
XLSR-SLS & ASV19 & 0.56 & 0.02 \\
XLSR-SLS & ASV5 & 27.00 & 0.62 \\
XLSR-SLS & AffectDF & 64.83 & 0.99 \\
\midrule
XLSR-Mamba & ASV19 & 0.20 & 0.01 \\
XLSR-Mamba & ASV5 & 13.65 & 0.30 \\
XLSR-Mamba & AffectDF & 46.47 & 0.99 \\
\midrule
ProSDD & ASV19 & 0.42 & 0.01 \\
ProSDD & ASV5 & 19.04 & 0.42 \\
ProSDD & AffectDF & 58.15 & 0.99 \\
\midrule
Voxtral & Inference & 46.06 & 0.93 \\
Voxtral & ASV19 & 3.05 & 0.09 \\
Voxtral & AffectDF & 19.52 & 0.63 \\
\bottomrule
\end{tabular}
\label{tab:tdcf_results}
\end{table}

\section{Effectiveness of Low-level Differential Cues}
\label{sec:low_level_sanity}

To examine whether the real/fake separation in AffectDF can be explained by simple low-level shortcuts, we perform an additional low-level differential cue analysis. We train a logistic-regression classifier using 8 low-level signal/channel features: RMS mean, RMS standard deviation, peak amplitude, clipping ratio, DC offset, zero-crossing rate, spectral rolloff, and high-frequency energy ratio. The classifier is trained on the AffectDF training set and evaluated on the AffectDF test set and ASVspoof2019 evaluation set. Results are reported in Table~\ref{tab:low_level_sanity}.

\begin{table}[t]
\centering
\small
\caption{
Evaluation of low-level differential cues using logistic regression trained on AffectDF.
}
\setlength{\tabcolsep}{5pt}
\renewcommand{\arraystretch}{1.15}
\begin{tabular}{llc}
\toprule
Train Data & Test Data & EER \\
\midrule
AffectDF train & AffectDF test & 53.16 \\
AffectDF train & ASVspoof2019 eval & 33.24 \\
\bottomrule
\end{tabular}
\label{tab:low_level_sanity}
\end{table}

As shown in Table~\ref{tab:low_level_sanity}, the low-level logistic-regression classifier obtains an EER of 53.16\% on the AffectDF test set, indicating that the selected low-level signal/channel features alone are not sufficient for separating real and fake samples in AffectDF. The same AffectDF-trained classifier performs better on ASVspoof2019, suggesting that these features can be informative when low-level real/fake differences are present. While this analysis considers only a limited set of low-level cues, the results suggest that AffectDF is less easily separated by such cues and requires models that capture more generalizable spoof-relevant representations.

\begin{figure*}[t]
    \centering
    \includegraphics[width=0.85\textwidth]{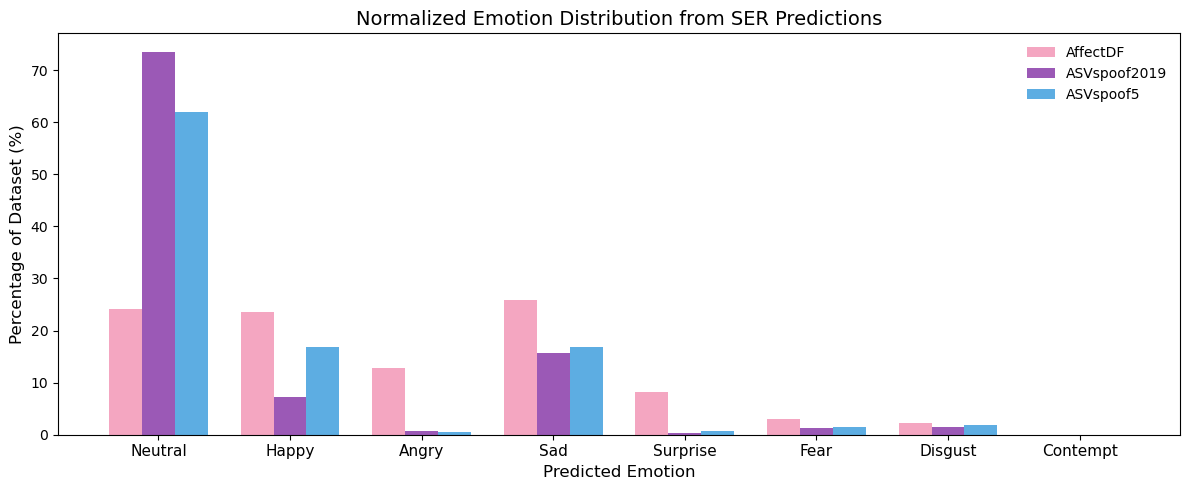}
    \caption{Normalized emotion distribution from Emotion2Vec+-large predictions across AffectDF and conventional SDD benchmarks.}
    \label{fig:ser_pred}
\end{figure*}

\begin{figure}[t]
    \centering
    \includegraphics[width=\columnwidth]{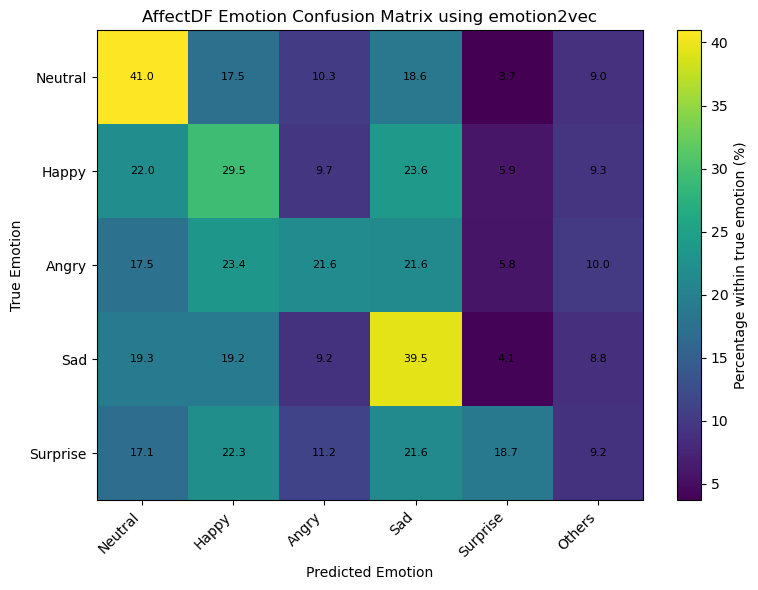}
    \caption{Emotion confusion matrix for AffectDF using Emotion2Vec+-large predictions.}
    \label{fig:ser_confusion}
\end{figure}

\section{The MSP-Podcast Corpus vs. the ESD Dataset}
\label{sec:MSPvsESD}
The difference between MSP-Podcast and ESD is more than simply spontaneous versus acted emotional speech. ESD is a controlled, parallel emotional speech corpus recorded in a studio-like environment, with 350 parallel utterances spoken by 10 native English across five emotion categories: neutral, happiness, anger, sadness, and surprise~\cite{ESD}. In contrast, MSP-Podcast is collected from naturally occurring podcast recordings, covering diverse speakers, topics, recording conditions, and conversational styles, with emotion labels obtained through perceptual annotation~\cite{Busso_202x}. Therefore, the ESD--MSP difference also reflects read versus natural speech, controlled versus in-the-wild recording conditions, parallel versus non-parallel utterances, and differences in speaker population and speaking style. Despite these corpus-level variations, our analysis focuses on the broader acted versus spontaneous distinction.

\subsection{Sample Size and Duration Statistics}
\label{sec:sample_duration_stats}

Table~\ref{tab:sample_duration_stats} reports the sample counts and average utterance duration for the acted and spontaneous attack subsets used in the acted-versus-spontaneous analysis. The acted subsets contain 6,000 samples per attack, while the spontaneous subsets contain approximately 4,100 samples per attack. The spontaneous samples are longer on average than the acted samples, so we report duration statistics alongside sample counts for this comparison. Conventional and SSL-based SDD systems use model-specific padding or cropping during training and evaluation, while LALM-based systems follow their corresponding audio-input pipelines, as described in Appendix~\ref{sec:model_training_details}.

\begin{table}[t]
\centering
\small
\caption{
Sample counts and average duration for acted and spontaneous attack subsets.
}
\setlength{\tabcolsep}{5pt}
\renewcommand{\arraystretch}{1.1}
\begin{tabular}{llcc}
\toprule
Category & Attack ID & Samples & Avg. Duration (s) \\
\midrule
Acted & A13 & 6,000 & 3.33 \\
Acted & A18 & 6,000 & 2.78 \\
Acted & A20 & 6,000 & 3.32 \\
Spontaneous & A14 & 4,108 & 6.98 \\
Spontaneous & A19 & 4,107 & 5.75 \\
Spontaneous & A21 & 4,107 & 5.72 \\
\bottomrule
\end{tabular}
\label{tab:sample_duration_stats}
\end{table}

\section{Automatic Evaluation of Emotion Expressiveness of AffectDF}
\label{sec:appendix_SER}

We further analyze the emotional characteristics of AffectDF and existing SDD benchmarks using speech emotion recognition (SER). Since the datasets differ substantially in size, Figure~\ref{fig:ser_pred} reports normalized emotion distributions as percentages rather than raw sample counts. We use the emotion2vec+large~\cite{emotion2vec} emotion classification model for this analysis. Although this model predicts seven emotion categories plus an ``unknown'' label, rather than the five categories used in AffectDF, it provides a consistent automatic measure of emotional expressiveness across datasets. We note that emotion2vec+ is fine-tuned using EmoBox~\cite{ma24b_interspeech}, which includes datasets such as MSP-Podcast and ESD; therefore, the SER results should be interpreted as an approximate expressiveness estimate rather than ground-truth emotion labels. Nevertheless, the normalized distributions show that AffectDF contains substantially stronger emotional coverage than conventional SDD benchmarks, supporting its role as an emotionally expressive benchmark for speech deepfake detection.

Figure~\ref{fig:ser_confusion} further reports the confusion matrix between AffectDF emotion labels and emotion2vec+ predictions. The automatic SER predictions do not always match the original emotion labels. However, the predictions are not dominated by the neutral class; instead, they are distributed across multiple emotional categories. This indicates that AffectDF contains meaningful emotional variability.

\section{Computational Details}
\label{sec:comp_deets}
All experiments were conducted on a shared GPU cluster using NVIDIA A100, H100, and H100-NVL GPUs, with access to up to eight GPUs of each type. Each conventional and SSL-based SDD model was trained under three training settings, using one A100 GPU per run. For LALM-based inference, Voxtral required one H100 GPU per dataset, while Qwen-2.5-Omni and Qwen-3.0-Omni required approximately two to three H100-NVL GPUs per evaluation. For synthetic speech generation, each generation model was run using one to two A100 GPUs, depending on the model size and inference requirements.

\section{Data License and Intended Use}
Data License and Intended Use. AffectDF is released for research and non-commercial use only under a CC BY-NC 4.0 license. The release includes generated audio and protocol files, subject to the licenses and terms of the original source corpora and generation models. The dataset is intended only for speech deepfake detection research and must not be used for impersonation, deception, biometric spoofing, or other harmful applications.

\section{Generative AI Use Disclosure}
\label{sec:ai}
Generative AI tools were employed solely for language polishing of text written by the authors. These tools were not used to generate scientific content, results, experimental designs, analyses, or conclusions. All authors are responsible for the full content of this paper and consent to its submission.
\end{document}